\documentclass[a4paper,12pt]{article}
\usepackage{amssymb}
\usepackage{amsmath}
\usepackage{amstext}
\usepackage{amsthm}
\usepackage{xspace}

\newcommand{\R}{\mathbb R}

\newcommand{\eps}{\varepsilon}

\renewcommand{\Im}{\mathrm{Im}}
\renewcommand{\Re}{\mathrm{Re}}
\newtheorem{theorem}{Theorem}
\newtheorem{lemma}[theorem]{Lemma}
\newtheorem{prop}[theorem]{Proposition}
\newtheorem{cor}[theorem]{Corollary}
\theoremstyle{definition}

\newtheorem{rem}[theorem]{Remark}
\theoremstyle{remark}

\begin{document}

\title{The Nonlinear Schroedinger equation: solitons dynamics}

\author{Vieri Benci  \thanks{Dipartimento di Matematica Applicata,
Universit\`a degli Studi di Pisa, Via F. Buonarroti 1/c, Pisa,
ITALY. e-mail: \texttt{benci@dma.unipi.it,
ghimenti@mail.dm.unipi.it, a.micheletti@dma.unipi.it}} \and Marco
Ghimenti \addtocounter{footnote}{-1}\footnotemark \and Anna Maria
Micheletti\addtocounter{footnote}{-1}\footnotemark }
\maketitle

\begin{abstract}

\noindent In this paper we investigate the dynamics of solitons
occurring in the nonlinear Schroedinger equation when a parameter
$h\rightarrow 0.$

\noindent We prove that under suitable assumptions, the the
soliton approximately
follows the dynamics of a point particle, namely, the motion of its \textit{%
barycenter} $q_{h}(t)$ satisfies the equation%
\[
\ddot{q}_{h}(t)+\nabla V(q_{h}(t))=H_{h}(t)
\]%
where
\[
\sup_{t\in \mathbb{R}}|H_{h}(t)|\rightarrow
0\;\;\text{as}\;\;h\rightarrow 0.
\]

\

\



\end{abstract}

\tableofcontents

\section{Introduction}

Roughly speaking a solitary wave is a solution of a field equation whose
energy travels as a localized packet and which preserves this localization
in time.

By \textit{soliton} we mean an \textit{orbitally stable} solitary
wave so that it has a particle-like behavior (for the definition
of orbital stability we refer e.g. to \cite{hylo}, \cite{BBGM07},
\cite{CL82} etc.).

In this paper we will be concerned with the dynamics of solitons relative to
a class of nonlinear Schroedinger equations (NSE).

Let us consider the following Cauchy problem relative to the NSE:
\begin{equation}
ih\frac{\partial \psi }{\partial t}=-\frac{h^{2}}{2}\Delta \psi +\frac{1}{%
2h^{\alpha }}W^{\prime }\left( h^{\gamma }|\psi |\right) \frac{\psi }{|\psi |%
}  \label{ch}
\end{equation}

\begin{equation}
\psi \left( 0,x\right) =\frac{1}{h^{\gamma }}U\left( \frac{x-q_{0}}{h^{\beta
}}\right) e^{\frac{i}{h}\mathbf{v}\cdot x}  \label{id}
\end{equation}%
where
\begin{equation*}
\beta =1+\frac{\alpha -\gamma }{2}
\end{equation*}%
and $U:\mathbb{R}^{N}\rightarrow \mathbb{R}$, $N\geq 2$, is a
positive, radially symmetric solution of the static nonlinear
Schroedinger equation
\begin{equation}
-\Delta U+W^{\prime }(U)=2\omega U  \label{eq}
\end{equation}%
with
\begin{equation}
\left\Vert U\right\Vert _{L^{2}}=\sigma  \label{sig}
\end{equation}%
Direct computations show that a solution of (\ref{ch}),(\ref{id}) is given
by
\begin{equation}
\psi \left( t,x\right) =\frac{1}{h^{\gamma }}U\left( \frac{x-q_{0}-\mathbf{v}%
t}{h^{\beta }}\right) e^{\frac{i}{h}\left( \mathbf{v}\cdot x-Et\right) }
\label{soli}
\end{equation}%
with%
\begin{equation*}
E=\frac{1}{2}\mathbf{v}^{2}+\frac{\omega }{h^{\alpha -\gamma }}
\end{equation*}%
Moreover if the problem (\ref{ch}),(\ref{id}) is well posed this is the
unique solution.

We can interprete this result saying that the \emph{barycenter} $q(t)$ of
the solution of (\ref{ch},\ref{id}) defined by
\begin{equation}
q(t)=\frac{\displaystyle\int_{\mathbb{R}^{N}}x|\psi (t,x)|^{2}dx}{%
\displaystyle\int_{\mathbb{R}^{N}}|\psi (t,x)|^{2}dx}  \label{bary}
\end{equation}%
satisfies the Cauchy problem
\begin{equation*}
\left\{
\begin{array}{c}
\ddot{q}=0 \\
q(0)=q_{0} \\
\dot{q}(0)=\mathbf{v}%
\end{array}%
\right.
\end{equation*}

The aim of this paper is to investigate what happens if the problem is
perturbed namely to investigate the problem
\begin{equation}
\left\{
\begin{array}{l}
\displaystyle ih\frac{\partial \psi }{\partial t}=-\frac{h^{2}}{2}\Delta
\psi +\frac{1}{2h^{\alpha }}W^{\prime }(h^{\gamma }|\psi |)\frac{\psi }{%
|\psi |}+V(x)\psi \\
\\
\displaystyle\psi \left( 0,x\right) =\varphi _{h}(x)%
\end{array}%
\right.  \tag{$P_{h}$}  \label{schr}
\end{equation}%
where
\begin{equation}
\varphi _{h}(x)=\left[ \frac{1}{h^{\gamma }}(U+w_{0})\left( \frac{x-q_{0}}{%
h^{\beta }}\right) \right] e^{\frac{i}{h}\mathbf{v}\cdot x}  \label{phih}
\end{equation}%
and $w_{0}$ is small, namely there is a constant $C$ such that%
\begin{eqnarray*}
\left\Vert w_{0}\right\Vert _{H^{1}} &\leq &Ch^{\alpha -\gamma } \\
\int_{\mathbb{R}^{N}}V(x)|w_{0}(x)|^{2}dx &<&Ch^{\alpha -\gamma }
\end{eqnarray*}%
Also we assume that%
\begin{equation*}
\left\Vert U+w_{0}\right\Vert _{L^{2}}=\left\Vert U\right\Vert
_{L^{2}}=\sigma
\end{equation*}

We make the following assumptions:

\begin{description}
\item (i) the problem (\ref{schr}) has a unique solution%
\begin{equation}
\psi \in C^{0}(\mathbb{R},H^{2}(\mathbb{R}^{N}))\cap C^{1}(\mathbb{R},
L^{2}(\mathbb{R}^{N}))  \label{gv}
\end{equation}
(sufficient conditions can be found in Kato \cite{Ka89}, Cazenave \cite{Ca03},
Ginibre-Velo \cite{GV79}; see Remark \ref{kato}).

\item (ii) $W:\mathbb{R}^{+}\rightarrow \mathbb{R}$ is a $C^{3}$ function
which satisfies the following assumptions:
\begin{equation}
W(0)=W^{\prime }(0)=W^{\prime \prime }(0)=0  \tag{$W_{0}$}  \label{W}
\end{equation}
\begin{equation}
|W^{\prime \prime }(s)|\leq c_{1}|s|^{q-2}+c_{2}|s|^{p-2}\text{ for some }
2<q\leq p<2^{\ast }.  \tag{$W_{1}$}  \label{Wp}
\end{equation}
\begin{equation}
W(s)\geq -c|s|^{\nu },\text{ }c\geq 0,\ 2<\nu <2+\frac{4}{N}
\text{ and }\ s\ \text{large}  \tag{$W_{2}$}  \label{W0}
\end{equation}
\begin{equation}
\exists s_{0}\in \mathbb{R}^{+}\text{ such that }W(s_{0})<0  \tag{$W_{3}$}
\label{W1}
\end{equation}

\item (iii) $V:\mathbb{R}^{N}\rightarrow \mathbb{R}$ is a $C^{2}$ function
which satisfies the following assumptions:
\begin{equation}
V(x)\geq 0;  \tag{$V_{0}$}  \label{V0}
\end{equation}%
\begin{equation}
|\nabla V(x)|\leq V(x)^{b}\text{ for }|x|>R_{1}>1,b\in (0,1);  \tag{$V_{1}$}
\label{Vinf*}
\end{equation}
\begin{equation}
V(x)\geq |x|^{a}\text{ for }|x|>R_{1}>1,a>1.  \tag{$V_{2}$}  \label{Vinf1}
\end{equation}
\end{description}

The main result of this paper is the following theorem:

\begin{theorem}
\label{teo1} Assume that (i), (ii) and (iii) hold and that%
\begin{equation}
\alpha >\gamma  \tag{\sc{crucial assumption}}
\end{equation}

Then the barycenter $q_{h}(t)$ of the solution of the problem (\ref{schr})
satisfies the following Cauchy problem:
\begin{equation*}
\left\{
\begin{array}{l}
\ddot{q}_{h}(t)+\nabla V(q_{h}(t))=H_{h}(t) \\
q_{h}(0)=q_{0} \\
\dot{q}_{h}(0)=\mathbf{v}%
\end{array}%
\right.
\end{equation*}%
where
\begin{equation*}
\sup_{t\in \mathbb{R}}|H_{h}(t)|\rightarrow 0\;\;\text{as}\;\;h\rightarrow 0
\end{equation*}
\end{theorem}

Let us discuss the set of our assumptions

\begin{rem}
\label{kato}About the assumption (i), we recall a result on the global
existence of solutions of the Cauchy problem (\ref{schr}) (see
\cite{Ca03,GV79,Ka89}). Assume (\ref{Wp}), (\ref{W0}) and (\ref{W1}) for $W$. Let
$D(A)$ (resp. $D(A^{1/2})$) denote the domain of the selfadjoint operator $A$
(resp. $A^{1/2}$) where
\begin{equation*}
A=-\Delta +V:L^{2}(\mathbb{R}^{N})\rightarrow L^{2}(\mathbb{R}^{N}).
\end{equation*}%
If $V\geq 0$, $V\in C^{2}$ and $|\partial ^{2}V|\in L^{\infty }$ and the
initial data $\psi (0,x)\in D(A^{1/2})$ then there exists the global
solution $\psi $ of (\ref{schr}) and

\begin{equation*}
\psi (t,x)\in C^{0}\left( \mathbb{R},D(A^{1/2})\right) \cap
C^{1}(\mathbb{R},H^{-1}(\mathbb{R}^{N})).
\end{equation*}
Furthermore, if $\psi (0,x)\in D(A)$ then
\begin{equation*}
\psi (t,x)\in C^{0}(\mathbb{R},D(A))\cap C^{1}(\mathbb{R},L^{2}(\mathbb{R}%
^{N})).
\end{equation*}
In this case, since $D(A)\subset H^{2}(\mathbb{R}^{N}),$ (i) is satisfied.
\end{rem}

\begin{rem}
The conditions (\ref{W}) and (\ref{V0}) are assumed for simplicity; in fact
they can be weakened as follows
\begin{equation*}
W^{\prime \prime }(0)=E_{0}
\end{equation*}
and%
\begin{equation*}
V(x)\geq E_{1}.
\end{equation*}
In fact, in the general case, the solution of the Schroedinger equation is
modified only by a phase factor.
\end{rem}

\begin{rem}
In \cite{BBGM07} the authors prove that if (ii) holds equation (\ref{ch})
admits orbitally stable solitary waves having the form (\ref{id}). In
particular the authors show that, under assumptions (\ref{Wp}), (\ref{W0})
and (\ref{W1}), for any $\sigma $ there exists a minimizer
$U(x)=U_{\sigma}(x)$ of the functional
\begin{equation*}
J(u)=\int \left( \frac{1}{2}\left\vert \nabla u\right\vert ^{2}+W(u)\right)
dx
\end{equation*}%
on the manifold $S_{\sigma }:=\{u\in H^{1},\ ||u||_{L^{2}}=\sigma \}$. Such
a minimizer satisfies eq.(\ref{eq}) where $2\omega $ is a Lagrange
multiplier. We will call \emph{ground state solution }a minimizer radially
symmetric around the origin.\emph{\ }We recall that by a well known result
of Gidas, Ni, and Nirenberg \cite{GNN81}, any positive solution of
eq.(\ref{eq}) is radially symmetric around some point.
\end{rem}

\begin{rem}
We set
\begin{equation*}
u_{h}(x)=h^{-\gamma }U\left( \frac{x}{h^{\beta }}\right)
\end{equation*}
where $U$ is a ground state solution. Now we establish a relation between
$\alpha ,\beta $ and $\gamma $ in order to have stationary solution of
(\ref{ch}) of the form $\psi (t,x)=u_{h}(x)e^{i\frac{\omega _{h}}{h}t}$, namely,
$u_{h}(x)\ $is a solution of the equation
\begin{equation*}
-h^{2}\Delta u_{h}+\frac{1}{h^{\alpha }}W^{\prime
}(h^{\gamma}u_{h})=2\omega _{h}u_{h}.
\end{equation*}
In fact, replacing $u_{h}$ by its explicit expression, we get
\begin{equation*}
-h^{2-\gamma }\Delta \left[ U\left( \frac{x}{h^{\beta }}\right) \right] +
\frac{1}{h^{\alpha }}W^{\prime }\left( U\left( \frac{x}{h^{\beta }}\right)
\right) =2\omega _{h}h^{-\gamma }U\left( \frac{x}{h^{\beta }}\right)
\end{equation*}
and hence, by rescaling the variable $x$,
\begin{equation*}
-h^{2-\gamma -2\beta +\alpha }\Delta U(x)+W^{\prime }\left( U(x)\right)
=2\omega _{h}h^{\alpha -\gamma }U\left( x\right) .
\end{equation*}
Thus, it is sufficient to take
\begin{equation}
\beta =1+\frac{\alpha -\gamma }{2}  \label{relfond}
\end{equation}
and%
\begin{equation}
\omega _{h}=\frac{\omega }{h^{\alpha -\gamma }}
\end{equation}
to obtain the claim. In the following we always assume (\ref{relfond}).
\end{rem}

\begin{rem}
The assumption (\ref{Vinf1}) is necessary if we want to identify
the position of the soliton with the barycenter (\ref{bary}). Let
us see why. Consider a soliton $\psi (x)$ and a perturbation
\begin{equation*}
\psi _{d}(x)=\psi (x)+\varphi \left( x-d\right) ,\ d\in \mathbb{R}^{N}
\end{equation*}
Even if $\varphi \left( x\right) \ll \psi (x),$ when $d$ is very large, the
``position" of $\psi (x)$ and the barycenter of $\psi _{d}(x)$ are far from
each other. In Lemma \ref{lemmabar5}, we shall prove that this situation
cannot occur provided that (\ref{Vinf1}) hold. In a paper in preparation, we
give a more involved notion of barycenter of the soliton and we will be able
to consider other situations.
\end{rem}

\begin{rem}
We will give a rough explanation of the assumption $\alpha >\gamma
$ which, in this approach to the problem, is crucial. In section
\ref{fio} we will show that the energy $E_{h}$ of a soliton $\psi
_{h}$ is composed by two parts: the internal energy $J_{h}$ and
the dynamical energy $G.$ The internal energy is a kind of binding
energy that prevents the soliton from splitting, while the
dynamical energy is related to the motion and it is composed of
potential and kinetic energy. As $h\rightarrow 0,$ we have that
(see section \ref{fio})
\begin{equation*}
J_{h}\left( \psi _{h}\right) \cong h^{N\beta -\alpha -\gamma }
\end{equation*}
and%
\begin{equation*}
G\left( \psi _{h}\right) \cong ||\psi _{h}||_{L^{2}}^{2}\cong h^{N\beta
-2\gamma }
\end{equation*}
Then, we have that
\begin{equation*}
\frac{G\left( \psi _{h}\right) }{J_{h}\left( \psi _{h}\right) }\cong
h^{\alpha -\gamma }
\end{equation*}
So the assumption $\alpha -\gamma >0$ implies that, for $h\ll 1,$
$G\left(\psi _{h}\right) \ll J_{h}\left( \psi _{h}\right)$, namely
the internal energy is bigger than the dynamical energy. This is
the fact that guarantees the existence and the stability of the
travelling soliton for any time.
\end{rem}

As far we know, this is the first paper in which there is a result
of type Th. \ref{teo1} for \emph{all} times $t \in \mathbb{R}$.
 However there are other results which compare the
motion of the soliton with the solution of the equation $\ddot
X(t)+\nabla V(X(t))=0$ for $t \in [0,T]$ for some constant $T <
\infty$.

Earlier results for pure power nonlinearity and bounded external
potential are in \cite{BJ00}. The authors have shown that if the
initial data is close to $U(\frac{x-q_0}{h})e^{i\frac{v_0\cdot
c}{h}}$ in a suitable sense then the solution $\psi_h(t,x)$ of
(\ref{schr}) satisfies for $t\in[0,T]$
\begin{equation}
\left\Vert\frac{1}{h^N}|\psi_h(t,x)|^2-\frac{1}{h^N}
\int_{\mathbb{R}^N}|\psi_h(t,x)|^2dx\delta_{X(t)}\right\Vert_{C^{1*}} \rightarrow0\text{ as }
h\rightarrow0.
\end{equation}
Here $C^{1*}$ is the dual of $C^1$ and $X(t)$ satisfies $\frac 12 \ddot
X(t)=\nabla V(X(t))$ with $X(0)=q_0$, $\dot X(0)=v_0$.

In related papers \cite{kera02} and \cite{kera06} there are slight
generalizations of the above result. Using a similar approach,
Marco Squassina \cite{squa} described the soliton dynamics in an
external magnetic potential.

In \cite{FGJS04} and \cite{FGJS06} the authors study the case of bounded
external potential $V$ respectively in $L^\infty$ or confining.

A result comparable with Theorem \ref{mainteoloc} is contained in \cite{FGJS06}.
The authors assume the existence of a stable ground state solution
with a null space non degeneracy condition of the equation
\begin{equation}
-\Delta \eta_\mu+\mu\eta_\mu+W^{\prime }(\eta_\mu)=0.
\end{equation}
The authors define a parameter $\varepsilon$ which depends on $\mu$ and on
other parameters of the problem. Under suitable assumptions they prove that
there exists $T>0$ such that, if the initial data $\psi^0(x)$ is very close
to $e^{ip_0\cdot(x-a_0)+i\gamma_0}\eta_{\mu_0}(x-a_0)$ the solution
$\psi(t,x)$ of problem $(P_1)$ with initial data $\psi^0$ is given by
\begin{equation}
\psi(t,x)=e^{ip(t)\cdot(x-a(t))+i\gamma(t)}\eta_{\mu(t)}(x-a(t))+w(t)
\end{equation}
with $||w||_{H^1}\leq \varepsilon$, $\dot p=-\nabla V(a)+o(\varepsilon^2)$,
$\dot a=2p+o(\varepsilon^2)$ with $0<t<\frac{T}{\varepsilon}$ for
$\varepsilon $ small.

The main differences with our result are the following. First of
all we do not have any limitation on the time $t$. Also, we have
an explicit estimate on $\ddot q$ (which roughly speaking
correspond to $\ddot a$). Our assumption on the nonlinearity $W$
are explicit (namely (\ref{Wp}), (\ref{W0}), (\ref{W1})) and we do
not require the null space condition which are, in general, not
easy to verify.

\subsection{Notations}

In the next we will use the following notations:
\begin{eqnarray*}
&&\Re (z),\Im (z)\text{ are the real and the imaginary part of }z \\
&&B(x_{0},\rho )=\{x\in \mathbb{R}^{N}\ :\ |x-x_{0}|\leq \rho \}\\
&&B(x_{0},\rho )^C= \mathbb{R}^{N}\smallsetminus B(x_{0},\rho )\\
&&S_{\sigma }=\{u\in H^{1}\ :\ ||u||_{L^{2}}=\sigma \} \\
&&J_{h}^{c}=\{u\in H^{1}\ :\ J_{h}(u)<c\} \\
&&U_{q}(x)=U(x-q) \\
&&\partial _{t}\psi =\frac{\partial }{\partial t}\psi \\
&&|\partial ^{\alpha }V(x)|=\sup_{i_{1},\dots ,i_{\alpha }}\left\vert
\frac{\partial ^{\alpha }V(x)}{\partial x_{i_{1}}\dots \partial x_{i_{\alpha }}}
\right\vert \text{ where }\alpha \in \mathbb{N},\ i_{1},\dots ,i_{\alpha}\in \{1,\dots ,N\} \\
&&I_{\sigma^2}=\inf_{ u\in H^1,\ \int u^2=\sigma^2}J(u)=m
\end{eqnarray*}

\section{General features of NSE}

Equation (\ref{schr}) is the Euler-Lagrange equation relative to the
Lagrangian density
\begin{equation}
\mathcal{L}=ih\partial _{t}\psi \overline{\psi }-\frac{h^{2}}{2}\left\vert
\nabla \psi \right\vert ^{2}-W_{h}(\psi )-V(x)\left\vert \psi \right\vert
^{2}  \label{lagr}
\end{equation}
where, in order to simplify the notation we have set
\begin{equation*}
W_{h}(\psi )=\frac{1}{h^{\alpha +\gamma }}W(h^{\gamma }\left\vert \psi
\right\vert )
\end{equation*}

Sometimes it is useful to write $\psi $ in polar form
\begin{equation}
\psi (t,x)=u(t,x)e^{iS(t,x)/h}.  \label{polar1}
\end{equation}
Thus the state of the system ${\psi }$ is uniquely defined by the couple of
variables $(u,S)$. Using these variables, the action $\mathcal{S=}\int
\mathcal{L}dxdt$ takes the form
\begin{equation}
\mathcal{S}(u,S)=-\int \left[ \frac{h^{2}}{2}\left\vert \nabla u\right\vert
^{2}+W_{h}(u)+\left( \partial _{t}S+\frac{1}{2}\left\vert \nabla
S\right\vert ^{2}+V(x)\right) u^{2}\right] dxdt  \label{polarAZ}
\end{equation}
and equation \ref{schr} becomes:

\begin{equation}
-\frac{h^{2}}{2}\Delta u+W_{h}^{\prime }(u)+\left( \partial _{t}S+
\frac{1}{2}
\left\vert \nabla S\right\vert ^{2}+V(x)\right) u=0  \label{Sh1}
\end{equation}

\begin{equation}
\partial _{t}\left( u^{2}\right) +\nabla \cdot \left( u^{2}\nabla S\right) =0
\label{Sh2}
\end{equation}

\subsection{The first integrals of NSE\label{fio}}

Noether's theorem states that any invariance for a one-parameter group of
the Lagrangian implies the existence of an integral of motion (see e.g.
\cite{Gelfand}).

Now we describe the first integrals which will be relevant for this paper,
namely the energy and the "hylenic charge".

\begin{description}
\item[Energy] The energy, by definition, is the quantity which is
preserved by the time invariance of the Lagrangian; it has the following
form
\begin{equation}
E_{h}(\psi )=\int \left[ \frac{h^{2}}{2}\left\vert \nabla \psi \right\vert
^{2}+W_{h}(\psi )+V(x)\left\vert \psi \right\vert ^{2}\right] dx
\end{equation}
Using (\ref{polar1}) we get:
\begin{equation}
E_{h}(\psi )=\int \left( \frac{h^{2}}{2}\left\vert \nabla u\right\vert
^{2}+W_{h}(u)\right) dx+\int \left( \frac{1}{2}\left\vert \nabla
S\right\vert ^{2}+V(x)\right) u^{2}dx  \label{Schenergy}
\end{equation}
Thus the energy has two components: the \textit{internal energy} (which,
sometimes, is also called \textit{binding energy})
\begin{equation}
J_{h}(u)=\int \left( \frac{h^{2}}{2}\left\vert \nabla u\right\vert
^{2}+W_{h}(u)\right) dx  \label{j}
\end{equation}

and the \textit{dynamical energy}
\begin{equation}
G(u,S)=\int \left( \frac{1}{2}\left\vert \nabla S\right\vert
^{2}+V(x)\right) u^{2}dx  \label{g}
\end{equation}
which is composed by the \textit{kinetic energy} $\frac{1}{2}\int \left\vert
\nabla S\right\vert ^{2}u^{2}dx$ and the \textit{potential energy} $\int
V(x)u^{2}dx$.

By our assumptions, the internal energy is bounded from below and the
dynamical energy is positive.

\item[Hylenic charge] Following \cite{hylo} the \textit{hylenic
charge}, is defined as the quantity which is preserved by by the
invariance of the Lagrangian with respect to the action
\begin{equation*}
\psi \mapsto e^{i\theta }\psi .
\end{equation*}
For equation (\ref{schr}) the charge is nothing else but the $L^{2}$ norm,
namely:
\begin{equation*}
\mathcal{H}(\psi )=\int \left\vert \psi \right\vert ^{2}dx=\int u^{2}dx
\end{equation*}
\end{description}

Now we study the rescaling properties of the internal energy and the $L^{2}$
norm of a function $u(x)$ having the form
\begin{equation*}
u(x):=h^{-\gamma }v\left( \frac{x}{h^{\beta }}\right)
\end{equation*}
We have
\begin{equation*}
||u||_{L^{2}}^{2}=h^{-2\gamma }\int v\left( \frac{x}{h^{\beta }}\right)
^{2}dx=h^{N\beta -2\gamma }\int v\left( \xi \right) ^{2}d\xi =h^{N\beta
-2\gamma }||v||_{L^{2}}^{2}.
\end{equation*}
and

\begin{eqnarray*}
J_{h}(u) &=&\int \frac{h^{2}}{2}|\nabla u|^{2}+\frac{1}{h^{\alpha +\gamma }}%
W(h^{\gamma }u)dx= \\
&=&\int \frac{h^{2-2\gamma }}{2}\left\vert \nabla _{x}v\left( \frac{x}{%
h^{\beta }}\right) \right\vert ^{2}+\frac{1}{h^{\alpha +\gamma }}W\left(
v\left( \frac{x}{h^{\beta }}\right) \right) dx= \\
&=&\int \frac{h^{N\beta +2-2\gamma -2\beta }}{2}\left\vert \nabla _{\xi
}v\left( \xi \right) \right\vert ^{2}+h^{N\beta -\alpha -\gamma }W\left(
v\left( \xi \right) \right) d\xi = \\
&=&h^{N\beta -\alpha -\gamma }\int \frac{1}{2}\left\vert \nabla _{\xi
}v\left( \xi \right) \right\vert ^{2}+W\left( v\left( \xi \right) \right)
d\xi =h^{N\beta -\alpha -\gamma }J_{1}(v),
\end{eqnarray*}
using the fundamental relation (\ref{relfond}).

\begin{rem}
If we choose $N\beta -2\gamma =0,$ the $L^{2}$ norm does not change by
rescaling. This implies that the dynamical energy $G,$ for $h$ small,
changes very little.
\end{rem}

\subsection{The swarm interpretation}

In this section we will suppose that the soliton is composed by a swarm of
particles which follow the laws of classical dynamics given by the
Hamilton-Jacobi equation. This interpretation will permit us to give an
heuristic proof of the main result.

First of all let us write NSE with the usual physical constants $m$ and
$\hslash$:
\begin{equation*}
i\hslash \frac{\partial \psi }{\partial t}=-\frac{\hslash ^{2}}{2m}\Delta
\psi +\frac{1}{2}W_{\hslash }^{\prime }(\psi )+V(x)\psi.
\end{equation*}
Here $m$ has the dimension of \textit{mass} and $\hslash$, the Plank
constant, has the dimension of \textit{action}.

In this case equations (\ref{Sh1}) and (\ref{Sh2}) become:

\begin{equation}
-\frac{\hslash ^{2}}{2m}\Delta u+\frac{1}{2}W_{\hslash }^{\prime }(u)+
\left(\partial _{t}S+\frac{1}{2m}\left\vert \nabla S\right\vert ^{2}+V(x)\right)
u=0  \label{Sh1c}
\end{equation}

\begin{equation}
\partial _{t}\left( u^{2}\right) +\nabla \cdot
\left( u^{2}\frac{\nabla S}{m}\right) =0  \label{Sh2c}
\end{equation}

The second equation allows us to interprete the matter field to be a fluid
composed by particles whose density is given by
\begin{equation*}
\rho _{\mathcal{H}}=u^{2}
\end{equation*}
and which move in the velocity field
\begin{equation}
\mathbf{v}=\frac{\nabla S}{m}.  \label{linda}
\end{equation}
So equation (\ref{Sh2c}) becomes the continuity equation:
\begin{equation*}
\partial _{t}\rho _{\mathcal{H}}+\nabla \cdot \left( \rho _{\mathcal{H}}
\mathbf{v}\right) =0.
\end{equation*}

If
\begin{equation}
-\frac{\hslash ^{2}}{2m}\Delta u+\frac{1}{2}W_{\hslash }^{\prime }(u)\ll u,
\label{rosaS}
\end{equation}
equation (\ref{Sh1c}) can be approximated by the eikonal equation
\begin{equation}
\partial _{t}S+\frac{1}{2m}\left\vert \nabla S\right\vert ^{2}+V(x)=0.
\label{hjS}
\end{equation}
This is the Hamilton-Jacobi equation of a particle of mass $m$ in a
potential field $V$.

If we do not assume (\ref{rosaS}), equation (\ref{hjS}) needs to be replaced
by
\begin{equation}
\partial _{t}S+\frac{1}{2m}\left\vert \nabla S\right\vert ^{2}+V+Q(u)=0
\label{hjqS}
\end{equation}
with
\begin{equation*}
Q(u)=\frac{-\left( \hslash ^{2}/m\right) \Delta u+
W_{\hslash }^{\prime }(u)}{2u}.
\end{equation*}

The term $Q(u)$ can be regarded as a field describing a sort of interaction
between particles.

Given a solution $S(t,x)$ of the Hamilton-Jacobi equation, the motion of the
particles is determined by eq.(\ref{linda}).

\subsection{An heuristic proof}

In this section we present an heuristic proof of the main result.
This proof is not at all rigorous, but it helps to understand the
underlying Physics.

\

If we interpret $\rho _{\mathcal{H}}=u^{2}$ as the density of
particles then
\begin{equation*}
\mathcal{H=}\int \rho _{\mathcal{H}}dx
\end{equation*}%
is the total number of particles. By (\ref{hjqS}), each of these particle
moves as a classical particle of mass $m$ and hence, we can apply to the
laws of classical dynamics. In particular the center of mass defined in
(\ref{bary}) takes the following form:
\begin{equation}
q(t)=\frac{\int xm\rho _{\mathcal{H}}dx}{\int m\rho _{\mathcal{H}}dx}=
\frac{\int x\rho _{\mathcal{H}}dx}{\int \rho _{\mathcal{H}}dx}.  \label{aiace}
\end{equation}
The motion of the barycenter is not affected by the interaction between
particles (namely by the term (\ref{hjqS})), but only by the external
forces, namely by $\nabla V.$ Thus the global external force acting on the
swarm of particles is given by
\begin{equation}
\overrightarrow{F}=-\int \nabla V(x)\rho _{\mathcal{H}}dx.  \label{ulisse}
\end{equation}
Thus the motion of the center of mass $q$ follows the Newton law
\begin{equation}
\overrightarrow{F}=M{\ddot{q}},  \label{tersite}
\end{equation}
where $M=\int m\rho _{\mathcal{H}}dx$ is the total mass of the swarm; thus
by (\ref{aiace}), (\ref{ulisse}) and (\ref{tersite}), we get
\begin{equation*}
{\ddot{q}}(t)=-\frac{\int \nabla V\rho _{\mathcal{H}}dx}
{m\int \rho _{\mathcal{H}}dx}=-\frac{\int \nabla Vu^{2}dx}{m\int u^{2}dx}.
\end{equation*}

If we assume that the $u(t,x)$ and hence $\rho _{\mathcal{H}}(t,x)$ is
concentrated in the point $q(t),$ we have that
\begin{equation*}
\int \nabla Vu^{2}dx\cong \nabla V\left( q(t)\right) \int u^{2}dx
\end{equation*}
and so, we get
\begin{equation*}
m{\ddot{q}}(t)\cong -\nabla V\left( q(t)\right).
\end{equation*}

Notice that the equation $m{\ddot{q}}(t)=-\nabla V\left( q(t)\right) $ is
the Newtonian form of the Hamilton-Jacobi equation (\ref{hjS}).

\section{Preliminary results}

In this section we prove two results which are the base of the main theorem.
They have some interest in themselves and require less assumptions than the
final theorem.

\subsection{Existence and dynamics of barycenter}

We recall the definition of barycenter of $\psi$

\begin{equation}
q_{h}(t)=\frac{\displaystyle\int_{\mathbb{R}^{N}}x|\psi (t,x)|^{2}dx}
{\displaystyle\int_{\mathbb{R}^{N}}|\psi (t,x)|^{2}dx}  \label{eqbar}
\end{equation}

The barycenter is not well defined for all the functions $\psi \in H^{1}
(\mathbb{R}^{N})$. Thus we need the following result:

\begin{theorem}
\label{din} Let $\psi (t,x)$ be a global solution of (\ref{schr}) such that
$\psi (t,x)\in C(\mathbb{R},H^1(\mathbb{R}^N))\cap C^1(\mathbb{R},H^{-1}
(\mathbb{R}^N))$ with initial data $\psi (0,x)$ such that
\begin{equation*}
\displaystyle \int_{\mathbb{R}^N} |x||\psi (0,x)|^2dx<+\infty.
\end{equation*}

Then the map $q_{h}(t):\mathbb{R}\rightarrow \mathbb{R}^{N},$ given by
(\ref{eqbar}) is well defined.
\end{theorem}

\textbf{Proof. }We show that $|\cdot |^{1/2}|\psi (t,\cdot )|\in L^{2}
(\mathbb{R}^{N})$ for any $t$, using a regularization argument.

We set
\begin{equation*}
k_{\varepsilon }(t)=\int_{\mathbb{R}^{N}}e^{-2\varepsilon |x|}|x||\psi
(t,x)|^{2}dx.
\end{equation*}
Since $\psi $ is a solution of (\ref{schr}), we have
\begin{eqnarray*}
k_{\varepsilon }^{\prime }(t) &=&\int_{\mathbb{R}^{N}}e^{-2\varepsilon
|x|}|x|[\partial _{t}\psi \bar{\psi}-\psi \partial _{t}\bar{\psi}]=2\Im
\left( \int i|x|\partial _{t}\psi \bar{\psi}e^{-2\varepsilon |x|}\right) = \\
&=&h\Im \left( \int \nabla \psi \nabla (|x|\bar{\psi}e^{-2\varepsilon
|x|})\right) +\frac{1}{h}\Im \left( \int 2|x|V|\psi |^{2}e^{-2\varepsilon
|x|}\right) + \\
&&+\frac{1}{h}\Im \left( \int \frac{|x|}{h^{\alpha }}W^{\prime -2\varepsilon
|x|}\right) = \\
&=&h\Im \left( \int \nabla \psi \nabla \bar{\psi}(|x|e^{-2\varepsilon
|x|})\right) +h\Im \left( \int \bar{\psi}\nabla \psi \nabla
(|x|e^{-2\varepsilon |x|})\right) = \\
&=&h\Im \left( \int \bar{\psi}\nabla \psi \cdot \frac{x}{|x|}
e^{-2\varepsilon |x|}(1-2\varepsilon |x|)\right) ,
\end{eqnarray*}
so we have
\begin{equation}
|k_{\varepsilon }^{\prime }(t)|\leq \int_{\mathbb{R}^{N}}|\bar{\psi}||\nabla
\psi |\leq ||\psi (t,\cdot )||_{L^{2}}||\nabla \psi (t,\cdot )||_{L^{2}},
\label{eqkprimo}
\end{equation}
then by (\ref{eqkprimo}) we get
\begin{eqnarray*}
k_{\varepsilon }(t) &=&k_{\varepsilon }(0)+h\Im \left( \int_{0}^{t}\int \bar{%
\psi}\nabla \psi \cdot \frac{x}{|x|}e^{-2\varepsilon |x|}(1-2\varepsilon
|x|)dxdt\right) \leq \\
&\leq &||\sqrt{|x|}\psi (0,x)||_{L^{2}}^{2}+\int_{0}^{t}||\psi (t,\cdot
)||_{L^{2}}||\nabla \psi (t,\cdot )||_{L^{2}}dt.
\end{eqnarray*}
By Fatou Lemma, when $\varepsilon \rightarrow 0$ we get $|\cdot |^{1/2}|\psi
(t,\cdot )\in L^{2}(\mathbb{R}^{N})$ for any $t\geq 0$. So the map
$q(t):[0,\infty )\rightarrow \mathbb{R}^{N}$ is well defined.

$\square $

\begin{theorem}
\label{don}The map $q_{h}(t):\mathbb{R}\rightarrow \mathbb{R}^{N},$ given by
(\ref{eqbar}) is $C^{1}$ and
\begin{equation}
\dot{q}_{h}(t)=\frac{\displaystyle\Im \left( h\int_{\mathbb{R}^{N}}\bar{\psi}(t,x)
\nabla \psi (t,x)dx\right) }{||\psi (t,x)||_{L^{2}}^{2}}.  \label{pina}
\end{equation}

Moreover if $\psi (t,x)\in C(\mathbb{R},H^{2}(\mathbb{R}^{N}))\cap C^{1}
(\mathbb{R},L^{2}(\mathbb{R}^{N}))$ then $q_{h}(t)$ is $C^{2}$ and
\begin{equation}
\ddot{q}_{h}(t)=\frac{\displaystyle\int_{\mathbb{R}^{N}}V(x)\nabla |\psi
(t,x)|^{2}dx}{||\psi (t,x)||_{L^{2}}^{2}}.  \label{pino}
\end{equation}
\end{theorem}

\textbf{Proof.} We have
\begin{equation*}
\dot{q}_{h}(t)=\frac{\displaystyle h\Im \left( \int_{\mathbb{R}^{N}}\bar{\psi}
\nabla \psi \right) }{\displaystyle||\psi (t,x)||_{L^{2}}^{2}}
\end{equation*}

We use the same regularization argument of Th. \ref{din}. We set
\begin{equation*}
K_{\varepsilon }^{i}(t)=\int_{\mathbb{R}^{N}}e^{-2\varepsilon |x|}x_{i}|\psi|^{2}dx
\end{equation*}
and again we find in the same way that
\begin{eqnarray*}
\frac{d}{dt}K_{\varepsilon }^{i}(t) &=&h\Im \left( \int \bar{\psi}\nabla
\psi \cdot e_{i}\ e^{-2\varepsilon |x|}\right) - \\
&&-h\Im \left( \int \bar{\psi}\nabla \psi \cdot \frac{x}{|x|}2\varepsilon
x_{i}\ e^{-2\varepsilon |x|}\right)
\end{eqnarray*}

where $e_i$ is the $i$-th vector of the canonical base of $\mathbb{R}^N$.
So, there exists a constant $c>0$ such that
\begin{equation*}
\left|\frac{d}{dt}K^i_\varepsilon(t)\right|\leq c||\psi
(t,\cdot)||_{L^2}||\nabla \psi (t,\cdot)||_{L^2},
\end{equation*}
Then we have that

\begin{equation*}
K^i_\varepsilon(t)=K^i_\varepsilon(0)+h\Im\left(\int_0^t \int \bar\psi
\nabla\psi \cdot e_i\ e^{-2\varepsilon |x|}- \int \bar\psi \nabla\psi \cdot
\frac{x}{|x|}2\varepsilon x_i\ e^{-2\varepsilon |x|} \right).
\end{equation*}

Using the dominated convergence theorem, when $\varepsilon \rightarrow 0$ we
have
\begin{equation*}
\int x_{i}|\psi (t,x)|^{2}dx=\int x_{i}|\psi (0,x)|^{2}dx+\int_{0}^{t}h\Im
\left( \int \bar{\psi}\nabla \psi \cdot e_{i}\right) dt,
\end{equation*}%
so, for all $i$ we have
\begin{equation}
\frac{d}{dt}\int_{\mathbb{R}^{N}}x_{i}|\psi |^{2}dx=h\Im \left( \int \bar{%
\psi}\nabla \psi \cdot e_{i}\right) .
\end{equation}%
This proves the first part of the theorem.

Next we prove that $\ddot{q}(t)=\frac{\displaystyle\int_{\mathbb{R}^{N}}V(x)\nabla |\psi (t,x)|^{2}dx}
{\displaystyle||\psi (t,x)||_{L^{2}}^{2}}$
under the supplementary assumption $\psi \in C^{1}(\mathbb{R},H^{1})$.

By this assumption we have that $\bar \psi (t,x)\nabla\psi (t,x)\in
C^1(\mathbb{R},L^1(\mathbb{R}^N))$. Thus
\begin{multline*}
\ddot q_h(t)=\frac{\displaystyle h\int_{\mathbb{R}^N} \Im(\partial_t[\bar
\psi (t,x)\nabla\psi (t,x)])dx} {||\psi(t,x)||_{L^{2}}^{2}}= \\
=\frac{\displaystyle h\int_{\mathbb{R}^N} \Im(\partial_t\bar \psi
(t,x)\nabla\psi (t,x)+ \bar \psi (t,x)\partial_t\nabla\psi (t,x))dx}
{||\psi(t,x)||_{L^{2}}^{2}}= \\
=\frac{\displaystyle h\int_{\mathbb{R}^N} \Im(\partial_t\bar \psi
(t,x)\nabla\psi (t,x)+ \bar \psi (t,x)\nabla\partial_t\psi (t,x))dx}
{||\psi(t,x)||_{L^{2}}^{2}}= \\
=\frac{\displaystyle h\int_{\mathbb{R}^N} \Im(\partial_t\bar \psi
(t,x)\nabla\psi (t,x)- \nabla\bar \psi (t,x)\partial_t\psi (t,x))dx}
{||\psi(t,x)||_{L^{2}}^{2}}= \\
=\frac{\displaystyle 2\Re\left(\int_{\mathbb{R}^N} ih\partial_t\psi
(t,x)\nabla\bar\psi (t,x)dx\right)} {||\psi(t,x)||_{L^{2}}^{2}}= \\
=\frac{\displaystyle \Re\left(\int_{\mathbb{R}^N}\left[-h^2\Delta\psi
+2V\psi + \frac {1}{h^\alpha}W^{\prime }(|\psi |)\frac{\psi }{|\psi |}\right]
\nabla\bar\psi dx\right)} {||\psi(t,x)||_{L^{2}}^{2}}.
\end{multline*}

We have, for all $i=1,\dots,N$,
\begin{equation*}
\Re \left(\int_{\mathbb{R}^N} W^{\prime }(|\psi |)\frac{\psi }{|\psi |}
\partial_{x_i}\bar\psi \right)= \int_{\mathbb{R}^N}\partial_{x_i}W(|\psi
|)=0,
\end{equation*}
because $W(|\psi |)\in L^1(\mathbb{R}^N)$ and $W^{\prime }(|\psi
|)\partial_{x_i}\bar\psi \in L^1$ because $\psi (t,\cdot)\in H^{2}$.

In the same way we have
\begin{equation*}
\Re \left(\int_{\mathbb{R}^N}-\Delta\psi \partial_{x_i}\bar\psi \right)=
\int_{\mathbb{R}^N}\partial_{x_i}|\nabla\psi |^2=0.
\end{equation*}
Thus
\begin{equation}
\ddot q_h(t)=\frac{\displaystyle 2\Re\left(\int_{\mathbb{R}^N}V\psi
\nabla\bar\psi dx\right)} {||\psi(t,x)||_{L^{2}}^{2}}= \frac{\displaystyle
\Re\left(\int_{\mathbb{R}^N}V(x)\nabla|\psi (t,x)|^2dx\right)}
{||\psi(t,x)||_{L^{2}}^{2}}.
\end{equation}
We point out that $V|\psi |\in L^1$ because $\psi $ is a global solution
with $\psi \in H^{2}$, $\partial_t\psi \in L^2$.

\noindent\emph{Step 3.} Conclusion.

Let $\psi (t,x)\in C^0(\mathbb{R},H^2)\cap C^1(\mathbb{R},L^2)$. We define a
function $\gamma_\lambda(t,x)\in C^0(\mathbb{R},H^2)\cap C^1(\mathbb{R},H^1)$
as
\begin{equation}
\gamma_\lambda(t,x)=\int_{\mathbb{R}^N}\varphi_\lambda(x-\xi)\psi (t,\xi)d
\xi
\end{equation}
where $\varphi(\xi)=\varphi(|\xi|)$ is a positive smooth function
with compact support in $|\xi|<\lambda$, with $\displaystyle
\int_{\mathbb{R}^N}\varphi_\lambda(\xi)d\xi=1$.

Fixed $t$ we have that
\begin{eqnarray*}
&&\gamma_\lambda(t,x)\rightarrow\psi (t,x)\text{ in }H^2(\mathbb{R}^N)
\text{ for }\lambda\rightarrow0; \\
&&\partial_t \gamma_\lambda(t,x)\rightarrow \partial_t\psi (t,x)\text{ in }
L^2(\mathbb{R}^N)\text{ for }\lambda\rightarrow0,
\end{eqnarray*}
and the convergence is uniform for every compact set in $\mathbb{R}$.

Furthermore, using that $\psi $ is a global solution in $C^0(\mathbb{R}
,H^2)\cap C^1(\mathbb{R},L^2)$ we have that $V(x)\psi (t,x)\in C^0(\mathbb{R}
,L^2)$ and that, fixed $t$
\begin{equation*}
V(x)\gamma_\lambda(t,x)\rightarrow V(x)\psi (t,x) \text{ in }L^2(\mathbb{R}^N)
\text{ for }\lambda\rightarrow0;
\end{equation*}
again, the convergence is uniform for every compact set in $\mathbb{R}$.

We have that $\gamma_\lambda(t,x)$ solve the following differential equation
\begin{multline*}
ih\frac{\partial \gamma_\lambda(t,x)}{\partial t}= \\
=-\frac{h^2}{2}\Delta\gamma_\lambda(t,x) +\frac 12 W^{\prime
}(|\gamma_\lambda(t,x)|)\frac{\gamma_\lambda(t,x)}{|\gamma_\lambda(t,x)|}+
V(x)\gamma_\lambda(t,x)+r(\gamma_\lambda(t,x))
\end{multline*}
where, %$r(\gamma_\lambda(t,x))\in C^0(\R,L^2)$ and,
$r(\gamma_\lambda(t,x))\rightarrow0$ in $L^2$, for all $t$, as
$\lambda\rightarrow0$, uniformly on every compact set in $\mathbb{R}$. Thus
we have, proceeding as in Step 3,
\begin{multline}
\frac{d}{dt}\int_{\mathbb{R}^N}\Im(\bar\gamma_\lambda(t,x)\gamma_\lambda(t,x))= \\
= \int_{\mathbb{R}^N}V(x)\nabla|\gamma_\lambda(t,x)|^2+
2\Re\big(r(\gamma_\lambda(t,x))\nabla\bar\gamma(t,x)\big)dx.
\end{multline}

We have
\begin{multline}  \label{eq46}
\int_{\mathbb{R}^N}\Im(\bar\gamma_\lambda(t,x)\gamma_\lambda(t,x))=
\int_{\mathbb{R}^N}\Im(\bar\gamma_\lambda(0,x)\gamma_\lambda(0,x))+ \\
+\int_0^t \int_{\mathbb{R}^N}V(x)\nabla|\gamma_\lambda(s,x)|^2+ 2\Re
\big(r(\gamma_\lambda(s,x))\nabla\bar\gamma(s,x)\big)dxds
\end{multline}
and, for all $s$,
\begin{equation*}
\int_{\mathbb{R}^N}V(x)\nabla|\gamma_\lambda(s,x)|^2+ 2\Re
\big(r(\gamma_\lambda(s,x))\nabla\bar\gamma(s,x)\big)dx\rightarrow
\int_{\mathbb{R}^N}V(x)\nabla|\psi (s,x)|^2
\end{equation*}
as $\lambda\rightarrow0$. Finally,
\begin{multline}  \label{eq47}
\int_{\mathbb{R}^N}V(x)\nabla|\gamma_\lambda(s,x)|^2+ 2\Re
\big(r(\gamma_\lambda(s,x))\nabla\bar\gamma(s,x)\big)dx\leq \\
\leq ||V(x)\gamma_\lambda(s,\cdot)||_{L^2}||\gamma_\lambda(s,\cdot)||_{H^1}+
||(r(\gamma_\lambda(s,\cdot))||_{L^2}||\gamma_\lambda(s,\cdot)||_{H^1}.
\end{multline}
And because $V(x)\gamma_\lambda(s,\cdot)\rightarrow V(x)\psi (s,\cdot)$ in
$L^2$, $r(\gamma_\lambda(s,\cdot)\rightarrow 0 $ in $L^2$ and
$\gamma_\lambda(s,\cdot)\rightarrow \psi (s,\cdot)$ in $H^1$ uniformly in $s$
on every compact set, we have that for some constant $C$
\begin{equation}
\sup_{s\in[0,t]}\int_{\mathbb{R}^N}V(x)\nabla|\gamma_\lambda(s,x)|^2+ 2\Re
\big(r(\gamma_\lambda(s,x))\nabla\bar\gamma(s,x)\big)dx\leq C.
\end{equation}
By (\ref{eq47}) we get
\begin{multline}  \label{eq51-52}
\int_0^t \int_{\mathbb{R}^N}V(x)\nabla|\gamma_\lambda(s,x)|^2dxds +\int_0^t
\int_{\mathbb{R}^N}2\Re\big(r(\gamma_\lambda(s,x))\nabla\bar\gamma(s,x)\big)
dxds\rightarrow \\
\int_0^t \int_{\mathbb{R}^N}V(x)\nabla|\psi (s,x)|^2dxds.
\end{multline}
Furthermore we know

\begin{eqnarray}
\label{eq49}
&&\int_{\mathbb{R}^{N}}\Im (\bar{\gamma}_{\lambda }(t,x)\gamma _{\lambda
}(t,x))\rightarrow \int_{\mathbb{R}^{N}}\Im (\bar{\psi}(t,x)\psi (t,x)); \\
\label{eq50}
&&\int_{\mathbb{R}^{N}}\Im (\bar{\gamma}_{\lambda }(0,x)\gamma _{\lambda
}(0,x))\rightarrow \int_{\mathbb{R}^{N}}\Im (\bar{\psi}(0,x)\psi (0,x)).
\end{eqnarray}
At this point by (\ref{eq46}), (\ref{eq51-52}), (\ref{eq49}) and (\ref{eq50})
we get
\begin{multline}
\int_{\mathbb{R}^{N}}\Im (\bar{\psi}(t,x)\psi (t,x))= \\
=\int_{\mathbb{R}^{N}}\Im (\bar{\psi}(0,x)\psi (0,x))+\int_{0}^{t}
\int_{\mathbb{R}^{N}}V(x)\nabla |\psi (s,x)|^{2}dxds
\end{multline}
that concludes the proof. %\end{proof}

We have the following corollary

\begin{cor}
\label{dincor}Assume (\ref{Vinf*}) and the assumptions of the previous
theorem; then
\begin{equation}
\ddot{q}_{h}(t)=-\frac{\displaystyle\int_{\mathbb{R}^{N}}\nabla V(x)|\psi
(t,x)|^{2}dx}{||\psi (t,x)||_{L^{2}}^{2}}.  \label{pinu}
\end{equation}
\end{cor}

\textbf{Proof.} By (\ref{Vinf*}), we have that%
\begin{eqnarray*}
\left\vert \int_{\mathbb{R}^{N}}\nabla V(x)|\psi (t,x)|^{2}dx\right\vert
&\leq &\int_{\mathbb{R}^{N}}V^{b}(x)|\psi (t,x)|^{2}dx \\
&\leq &C_1\int_{\mathbb{R}^{N}}V(x)|\psi (t,x)|^{2}dx \\
&\leq &C_{2}G\left( \psi \right) <+\infty
\end{eqnarray*}
where $G$ is the dynamical energy (\ref{g}). Thus, we can integrate by parts
and we have that
\begin{equation*}
\int_{\mathbb{R}^{N}}V(x)\nabla |\psi (t,x)|^{2}dx=
-\int_{\mathbb{R}^{N}}\nabla V(x)|\psi (t,x)|^{2}dx
\end{equation*}

$\square $

\begin{rem}
If we use the polar form (\ref{polar1}), (\ref{pina}) and (\ref{pinu}) take
the more meaningful form respectively:
\begin{equation*}
\dot{q}_{h}(t)=\frac{\displaystyle\int_{\mathbb{R}^{N}}\nabla S\ u^{2}dx}
{\displaystyle \int_{\mathbb{R}^{N}}u^{2}dx}
\end{equation*}
\begin{equation*}
\ddot{q}_{h}(t)=-\frac{\displaystyle\int_{\mathbb{R}^{N}}\nabla V(x)u^{2}dx}
{\displaystyle\int_{\mathbb{R}^{N}}u^{2}dx}
\end{equation*}
They can be interpreted as follows: $\dot{q}_{h}(t)$ is the
average momentum (remember (\ref{linda}) and that $m=1$);
$\ddot{q}_{h}(t)$ equals the average force, since
$\overrightarrow{F}\cong -\nabla V$ (see (\ref{ulisse})).
\end{rem}

\subsection{Concentration results}

In this section we prove a concentration property of the solution
of (\ref{schr}) with initial data (\ref{phih}); more exactly, we
prove that for \textit{any time} $t\in \mathbb{R}$, this solution
is a "bump" of radius less than some constant "$\hat{R}$". In
order to prove this result, it is sufficient to assume that
problem (\ref{schr}) admits global solutions $\psi (t,x)\in
C(\mathbb{R},H^{1}(\mathbb{R}^{N}))$ which satisfy the
conservation of the energy and of the $L^{2}$ norm, namely it is
not necessary to assume the regularity (\ref{gv}).

\bigskip

For some nonlinearities $W$, it is possible that the ground state solution is
not unique. In any case we have the following result:

\begin{prop}
\label{remstrauss} Let $U$ be a ground state solution of (\ref{eq}). Then,
for $|x|~>1$ and $N\geq 2$
\begin{equation*}
U(x)<\frac{C}{|x|^{\frac{N-1}{2}}}
\end{equation*}%
where $C$ is a constant which does not depend on $U$.
\end{prop}

\textbf{Proof. }By a well known inequality due to Strauss \cite{St77} we
have
\begin{equation}
0<U(x)\leq C_{N}\frac{||U||_{H^{1}}}{|x|^{\frac{N-1}{2}}}\text{ for }|x|\geq
\alpha _{N}\text{ a.e.}
\end{equation}
where $C_{N}$ and $\alpha _{N}$ depend only on the dimension $N$. Moreover
there exists a constant $C_{m}$, such that $||U||_{H^{1}}\leq C_{m}$ for any
$U$ minimizer of $\displaystyle\inf_{u\in S_{\sigma }}J(u)=m$. In fact we
have the following inequality
\begin{equation}
||u||_{L^{\nu }}\leq b_{\nu }||u||_{L^{2}}^{1-\frac{N}{2}+
\frac{N}{\nu }}||\nabla u||_{L^{2}}^{\frac{N}{2}-\frac{N}{\nu }}
\end{equation}
for some constant $b_{\nu }$. Then, by (\ref{sig})
\begin{equation}
||U||_{L^{\nu }}^{\nu }\leq b_{\nu }^{\nu }\sigma ^{\nu \left( {1-\frac{N}{2}%
+\frac{N}{\nu }}\right) }||\nabla U||_{L^{2}}^{\nu \frac{N}{2}-N}.
\label{remconc1}
\end{equation}
By assumption (\ref{W0}) and by (\ref{remconc1}) we have
\begin{eqnarray*}
m &=&J(U)\geq \int \frac{1}{2}|\nabla U|^{2}-c_{1}U^{2}-c_{2}U^{\nu }dx\geq
\\
&\geq &\frac{1}{2}\int |\nabla U|^{2}-c_{3}\left( \int |\nabla U|^{2}\right)
^{\nu \frac{N}{4}-\frac{N}{2}}-c_{1}\sigma ^{2}
\end{eqnarray*}
for some constant $c_{3}$. If $0<\nu \frac{N}{2}-N<2$, namely $2<\nu <2+%
\frac{4}{N},$ we have the claim.

$\square $

\begin{lemma}
\label{conc1} For any $\varepsilon >0$, there exists an $\hat{R}=\hat{R}(\varepsilon )$
and a $\delta =\delta (\varepsilon )$ such that, for any
$u\in J^{m+\delta }\cap S_{\sigma }$, we can find a point $\hat{q}=\hat{q}
(u)\in \mathbb{R}^{N}$ such that
\begin{equation}
\frac{1}{\sigma ^{2}}\int\limits_{\mathbb{R}^{N}\smallsetminus B(\hat{q},
\hat{R})}u^{2}(x)dx<\varepsilon .
\end{equation}
\end{lemma}

%\begin{proof}
\textbf{Proof. }Firstly we prove that for any $\varepsilon >0$,
there exists a $\delta $ such that, for any $u\in J^{m+\delta
}\cap S_{\sigma }$, we can find a point $\hat{q}=\hat{q}(u)\in
\mathbb{R}^{N}$ and a radial ground state solution $U$ such that
\begin{equation}
||u(x)-U(x-\hat{q})||_{H^{1}}\leq \varepsilon .
\end{equation}

We argue by contradiction. Suppose that there exists an $\varepsilon>0$ and
a sequence $\{u_n\}_n$ such that $||u_n||_{L^2}=\sigma$, $J(u_n)\rightarrow
m $ and, for any $q\in \mathbb{R}^N$ and for each $U$ ground state solution
it holds
\begin{equation}  \label{formconc1}
\varepsilon<||u(x)-U(x- q)||_{H^1}.
\end{equation}

By the Ekeland principle we can assume that $\{u_n\}$ is a Palais Smale
sequence for $J$ on $S_\sigma$, that is, there exists $\{\lambda_n\}$ such
that
\begin{equation}
-\Delta u_n+W^{\prime }(u_n)-\lambda_nu_n\rightarrow0\text{ as }
n\rightarrow\infty.
\end{equation}

By \cite[Proposition 11]{BBGM07} up to a subsequence we have that
$\lambda_n\rightarrow\bar\lambda<0$. So we get
\begin{eqnarray}
&&-\Delta u_n+W^{\prime }(u_n)-\bar\lambda u_n\rightarrow0; \\
&&J(u_n)-\bar\lambda\int u_n^2\rightarrow m-\bar\lambda\sigma^2.
\end{eqnarray}

As a consequence of the Concentration Compactness principle
\cite{Li84a,Li84b}, we can describe the behavior of this P.S. sequence. We use
the Splitting Lemma (see \cite{St84}, \cite{BBGM07}) and we get
\begin{eqnarray}
&&u_n=\sum_{j=1}^kU^j(x-q^j_n) + w_n\text{ with }w_n\rightarrow0 \text{ in }
H^1  \label{split1} \\
&&\sigma^2=\sum_{j=1}^k||U^j(x-q^j_n)||^2_{L^2}  \label{split2} \\
&&\sum_{j=1}^kJ(U^j(x-q^j_n))=m=I_{\sigma^2}  \label{split3}
\end{eqnarray}
where $U^j$ are solutions of $-\Delta u+W^{\prime }(u)=\bar\lambda u$ and
$q^j_n\in\mathbb{R}^N$.

Here $I_{\rho ^{2}}=\min_{||u||_{L^{2}}^{2}=\rho ^{2}}J(u)$. We recall (see
\cite{Li84a}) that for any $\mu \in (0,\rho )$ we have
\begin{equation}
I_{\rho ^{2}}<I_{\mu ^{2}}+I_{\rho ^{2}-\mu ^{2}}.  \label{concomp}
\end{equation}%
We verify that in (\ref{split1})-(\ref{split3}) it is $k=1$. We assume $k=2$.
Suppose that $||U^{1}||_{L^{2}}=\mu ^{2}<\sigma ^{2}$. Then, by (\ref%
{concomp}), we have a contradiction because
\begin{equation}
I_{\sigma ^{2}}<I_{\mu ^{2}}+I_{\sigma ^{2}-\mu ^{2}}\leq
J(U^{1})+J(U^{2})=I_{\sigma ^{2}}.
\end{equation}%
For the case $k>2$ we argue analogously.

Thus we have, up to subsequence,
\begin{equation}  \label{formconc2}
u_n(x)=U(x-q_n)+w_n\ \ \ \ w_n\rightarrow0 \text{ in }H^1
\end{equation}
for some $U$ radial ground state solution, and (\ref{formconc2}) contradicts
(\ref{formconc1}).

At this point, given $\varepsilon$, there exist a point $\hat q=\hat q(u)\in
\mathbb{R}^N$ and a radial ground state solution $U$ such that
\begin{equation}  \label{formconc3}
u(x)=U(x-\hat q)+w\text{ and } ||w||_{H^1}\leq C\varepsilon.
\end{equation}

Now, we choose $\hat R$ such that
\begin{equation}  \label{eqconc2}
\frac{1}{\sigma^2}\int\limits_{\mathbb{R}^N\smallsetminus B(0,\hat
R)}U^2(x)dx<C\varepsilon
\end{equation}
for all $U$ radial ground state solutions. This is possible because, if $U$
is a radial minimizer of $J(u)$ on $S_\sigma$, then, as showed in following
Remark \ref{remstrauss},
\begin{equation}  \label{formconc4}
U(x)\leq \frac{C(m,N)}{x^{\frac{N-1}2}}\text{ for }|x|>>1,\ N\geq2,
\end{equation}
the constant $C(m,N)$ depending only on the dimension $N$ and on
$\displaystyle m=\inf_{u\in S_\sigma}J(u)$.

We get
\begin{eqnarray}
\nonumber \frac{1}{\sigma^{2}}\int\limits_{B(\hat{q},\hat{R})^C}u^{2}(x)dx&<&\frac{1}{\sigma^{2}}
\int\limits_{B(\hat{q},\hat{R})^C}U^{2}(x-\hat{q})dx+\frac{1}{\sigma^{2}}
\int\limits_{B(\hat{q},\hat{R})^C}w^{2}+2wUdx
\\ \label{eqconc3}
&=&\frac{1}{\sigma ^{2}}\int\limits_{B(0,\hat{R})^C}U^{2}(x)dx+
\frac{1}{\sigma ^{2}}\int\limits_{B(\hat{q},\hat{R})^C}w^{2}+2wUdx.
\end{eqnarray}
By (\ref{formconc3}), (\ref{eqconc2}), (\ref{eqconc3}) we get the claim. We
notice also the $\hat{R}$ does not depend on $u,U$.

$\square $%\end{proof}

We can describe now the concentration properties of the solution of
(\ref{schr}).

\begin{lemma}
\label{lemmaconc} For any $\varepsilon>0$, there exists a
$\delta=\delta(\varepsilon)$ and a $\hat R=\hat R(\varepsilon)$ such that for
any $\psi (t,x)$ solution of (\ref{schr}) with $|h^{-\gamma}\psi (t,h^\beta
x)|\in J^{m+ \delta}\cap S_\sigma$ for all $t$ there exists a $\hat
q_h(t)\in \mathbb{R}^N$ for which
\begin{equation}
\frac{1}{\sigma^2h^{N\beta-2\gamma}}\int\limits_{\mathbb{R}^N\smallsetminus
B(\hat q_h(t),\hat Rh^\beta)}|\psi (t,x)|^2dx<\varepsilon.
\end{equation}
Here $\hat q_h(t)$ depends on $\varepsilon$ and $\psi (t,x)$.
\end{lemma}

\textbf{Proof. } Fixed $h$ and $t$, we set
$v(\xi)=|h^{-\gamma}\psi (t,h^\beta \xi)|$. So we have
\begin{eqnarray*}
m<J(v)\leq m+ \delta&\text{ and }&
||v||_{L^2}=\sigma.
\end{eqnarray*}
So, by Lemma \ref{conc1}, we have that there exist an $\hat R>0$ and a
$\bar q=\bar q(v)$ such that
\begin{equation}
\eps>\frac{1}{\sigma^2}\int\limits_{\R^N\smallsetminus B(\bar q,\hat R)}|v(\xi)|^2d\xi
\end{equation}
By a change of variable we obtain
\begin{equation}
\eps>\frac{1}{\sigma^2}\int\limits_{\R^N\smallsetminus B(\bar q,\hat R)}|v(\xi)|^2d\xi
=
\frac{1}{\sigma^2h^{N\beta-2\gamma}}
\int\limits_{\R^N\smallsetminus B(\hat q(t),\hat Rh^\beta)}|\psi (t,x)|^2dx,
\end{equation}
where $\hat q_h(t)$ depends on $\eps,h,t$  and $\psi $, while
$\hat R$ depends only by $\eps$.

$\square $

We give now some results about the concentration property of the solutions
$\psi (t,x)$ of the problem (\ref{schr}). Given $K>0$ $q\in \R^{N}$, $h>0$ we
call
\begin{equation}
B_{h}^{K,q}=\left\{
\begin{array}{c}
\psi (0,x)=u_{h}(0,x)e^{\frac{i}{h}S_{h}(0,x)}\text{ } \\
\text{with }u_{h}(0,x)=h^{-\gamma }\left[ (U+w)\left( \frac{x-q}{h^{\beta }}
\right) \right] \\
\\
U\text{ is a radial ground state solution} \\
\\
||U+w||_{L^{2}}=||U||_{L^{2}}=\sigma \text{ and }||w||_{H^{1}}<Kh^{\alpha
-\gamma } \\
\\
||\nabla S_{h}(0,x)||_{L^{\infty }}\leq K\text{ for all }h \\
\\
\int_{\mathbb{R}^{N}}V(x)u_{h}^{2}(0,x)dx\leq Kh^{N\beta -2\alpha }.
\end{array}
\right\}  \label{bkqh}
\end{equation}
the set of admissible initial data.

\begin{rem}
The condition $||w||_{H^1}\leq Kh^{\alpha-\gamma}$ can be weakened. Indeed
in the proof of the theorem we need $J(U+w)\leq m+Kh^{\alpha-\gamma}$, which
is implied by $||w||_{H^1}\leq Kh^{\alpha-\gamma}$. We prefer to refer to
the strongest but simpler hypotheses to simplify the statement of the main
theorem.
\end{rem}

\begin{rem}
In Theorem \ref{teo1} we assume $S_{h}(0,x)=v\cdot x$ which is more stronger
than $||\nabla S_{h}(0,x)||_{L^{\infty }}\leq K$ to simplify the statement
and for a better physical interpretation.
\end{rem}

Finally, we can prove the main result of this section.

\begin{theorem}
\label{teoconcinf} Assume $V\in L_{\text{loc}}^{\infty }$ and (\ref{V0}).
Fix $K>0$, $q\in \mathbb{R}^{N}$. Let $\alpha >\gamma $.

For all $\varepsilon >0$, there exists $\hat{R}>0$ and $h_{0}>0$ such that,
for any $\psi (t,x)$ solution of (\ref{schr}) with initial data $\psi
(0,x)\in B_{h}^{K,q}$ with $h<h_{0}$, and for any $t$, there exists
$\hat{q}_{h}(t)\in \mathbb{R}^{N}$ for which
\begin{equation}
\frac{1}{||\psi (t,x)||_{L^{2}}^{2}}
\int\limits_{\mathbb{R}^{N}\smallsetminus B(\hat{q}_{h}(t),\hat{R}h^{\beta })}
|\psi(t,x)|^{2}dx<\varepsilon .
\end{equation}
Here $\hat{q}_{h}(t)$ depends on $\psi (t,x)$.
\end{theorem}

\textbf{Proof. } By the conservation law, the energy $E_{h}(\psi
(t,x))$ is constant with respect to $t$. Then we have
\begin{eqnarray*}
E_{h}(\psi (t,x))&=&E_{h}(\psi (0,x))\\
&=&J_{h}(u_{h}(0,x))+\int_{\R^N}u_{h}^2(0,x)
\left[\frac{|\nabla S_{h}(0,x)|^2}2+V(x)\right]dx\\
&\leq&J_{h}(u_{h}(0,x))+\frac K2\sigma^2h^{N\beta-2\gamma}+Kh^{N\beta-2\gamma}\\
&=&h^{N\beta-\alpha-\gamma}J\left(U+w\right)+ h^{N\beta-2\gamma}C
\end{eqnarray*}
where $C$ is a suitable constant. Now, by rescaling, and using that
$\psi (0,x)\in B^{K,q}_h$, we obtain
\begin{eqnarray}
\nonumber E_{h}(\psi (t,x))
&=&h^{N\beta-\gamma-\alpha}J(U+w)+Ch^{N\beta-2\gamma}\\
&\leq&
h^{N\beta-\gamma-\alpha}(m+Kh^{\alpha-\gamma})+Ch^{N\beta-2\gamma}
\label{eq77bis}\\
\nonumber
&=&h^{N\beta-\gamma-\alpha}(m+Kh^{\alpha-\gamma}+Ch^{\alpha-\gamma})=
h^{N\beta-\gamma-\alpha}\big(m+h^{\alpha-\gamma}C_1\big)
\end{eqnarray}
where $C_1$ is a suitable constant. Thus
\begin{eqnarray}
\nonumber
J_{h}(u_{h_n}(t,x))&=&E_h(\psi (t,x))-G_h(\psi (t,x)\\
&=&E_h(\psi (t,x))-\int_{\R^N} \left[\frac{|\nabla
S_h(t,x)|^2}2+V(x)\right]u_h(t,x)^2dx\nonumber
\\
\label{Jh} &\leq&
h^{N\beta-\gamma-\alpha}\big(m+h^{\alpha-\gamma}C_1\big)
\end{eqnarray}
because $V\geq0$. By rescaling the inequality (\ref{Jh}) we get
\begin{equation}
J\big(h^{-\gamma}u_h(t,h^\beta x)\big)\leq m+h^{\alpha-\gamma}C_1
\end{equation}
So, if $\alpha>\gamma$, for $h$ small we can apply
Lemma \ref{lemmaconc} and we get the claim.

$\square $

\section{The final result}

\subsection{Barycenter and concentration point}

In this paragraph, we estimate the distance between the concentration point
and the barycenter of a solution $\psi (t,x)$ for a potential satisfying
hypothesis (\ref{V0}) and (\ref{Vinf1}).

Hereafter, fixed $K>0$, we assume that $\psi (t,x)$ is a global solution of
the Schr\"{o}dinger equation (\ref{schr}),
$\psi (t,x)\in C(\mathbb{R},H^{1})\cap C^{1}(\mathbb{R},H^{-1})$,
with initial data $\psi (0,x)\in
B_{h}^{K,q}$ with $B_{h}^{K,q}$ given by (\ref{bkqh}).

\begin{lemma}
\label{lemmabar1} There exists a constant $L>0$ such that
\begin{equation*}
0\leq \frac{1}{h^{N\beta -2\alpha }}\int_{\mathbb{R}^{N}}V(x)u_{h}^{2}(t,x)dx
\leq L\ \ \forall t\in \mathbb{R}.
\end{equation*}
\end{lemma}

%\begin{proof}
\textbf{Proof. }At first we notice that $||h^{-\gamma }u_{h}(t,h^{\beta}x)||_{L^{2}}^{2}=
||h^{-\gamma }u_{h}(0,h^{\beta}x)||_{L^{2}}^{2}=
||U+w||_{L^{2}}^{2}=\sigma ^{2}$. Thus
\begin{equation}
J_{h}(u_{h}(t,x))=h^{N\beta -\gamma -\beta }J(h^{-\gamma }u_{h}(t,h^{\beta}x))
\geq h^{N\beta -\gamma -\beta }m.
\end{equation}
By (\ref{eq77bis}), there exist a constant $L$ such that
\begin{equation}
E_{h}(\psi (t,x))\leq h^{N\beta -\gamma -\alpha }m+Lh^{N\beta -2\gamma }.
\end{equation}
Finally,
\begin{eqnarray*}
\int_{\mathbb{R}^{N}}V(x)u_{h}^{2}(t,x)dx &=&E_{h}(\psi
(t,x))-J_{h}(u_{h}(t,x))-\int_{\mathbb{R}^{N}}\frac{|\nabla S|^{2}}{2}
u_{h}^{2}(t,x)dx \\
&\leq &E_{h}(\psi (t,x))-J_{h}(u_{h}(t,x)) \\
&\leq &h^{N\beta -\gamma -\alpha }m+Lh^{N\beta -2\gamma }-h^{N\beta -\gamma
-\beta }m=Lh^{N\beta -2\gamma }
\end{eqnarray*}
that concludes the proof.

$\square $ %\end{proof}

\begin{rem}
\label{rembar2} By Lemma \ref{lemmabar1} we get, for any $R_{2}\geq R_{1}$
($R_{1}$ given in (\ref{Vinf1})) and for any $t\in \mathbb{R}$ the following
inequality
\begin{eqnarray}
L &\geq &\frac{1}{h^{N\beta -2\gamma }}\int_{|x|\geq
R_{2}}V(x)u_{h}^{2}(t,x)dx  \label{eqbar1} \\
&\geq &\frac{1}{h^{N\beta -2\gamma }}\int_{|x|\geq
R_{2}}|x|^{a}u_{h}^{2}(t,x)dx\geq \frac{R_{2}^{a-1}}{h^{N\beta -2\gamma }}
\int_{|x|\geq R_{2}}|x|u_{h}^{2}(t,x)dx  \notag
\end{eqnarray}
\end{rem}

\begin{lemma}
\label{lemmabar3} There exists a constant $K_{1}$ such that
\begin{equation*}
|q_{h}(t)|\leq K_{1}\text{ for }t\in \mathbb{R}.
\end{equation*}
\end{lemma}

By Lemma \ref{lemmabar1} and Remark \ref{rembar2} we have that
\begin{eqnarray*}
\left\vert \int_{\mathbb{R}^{N}}xu_{h}^{2}(t,x)dx\right\vert  &\leq
&\int_{|x|\geq R_{1}}|x|u_{h}^{2}(t,x)+\int_{|x|<R_{1}}|x|u_{h}^{2}(t,x) \\
&\leq &R_{1}\int_{\mathbb{R}^{N}}u_{h}^{2}(t,x)dx+\frac{L}{R_{1}^{a-1}}
h^{N\beta -2\gamma }.
\end{eqnarray*}
So, using the definition of $q_{h}(t)$ we have
\begin{equation}
|q_{h}(t)|\leq R_{1}+\frac{L}{R^{a-1}\sigma ^{2}}=K_{1},
\end{equation}
for some $K_{1}>0$. %

\begin{rem}
\label{rembar4} By the inequality (\ref{eqbar1}) in Remark \ref{rembar2}, we
have also that, for any $R_{2}\geq R_{1}$,
\begin{equation*}
\frac{\displaystyle\int_{|x|\geq R_{2}}u_{h}^{2}(t,x)dx}
{\displaystyle\int_{\mathbb{R}^{N}}u_{h}^{2}(t,x)dx}\leq
\frac{L}{\sigma ^{2}R_{2}^{a}} \text{ for all }t\in \mathbb{R}.
\end{equation*}
Hereafter, we always choose $R_{2}$ large enough to have
\begin{equation}
\frac{L}{\sigma ^{2}R_{2}^{a}}<\frac{1}{2}.
\end{equation}
\end{rem}

Now we show the boundedness of the concentration point $\hat{q}_{h}(t)$
defined in Lemma (\ref{lemmaconc}).

\begin{lemma}
\label{lemmabar5} Given $0<\varepsilon<1/2$, and $R_2$ as in the previous
remark.

We get

\begin{enumerate}
\item $\sup\limits_{t\in\mathbb{R}}|\hat q_h(t)|<R_2+\hat
R(\varepsilon)h^\beta<R_2+1$, for all $h<\bar h$ and $\delta<\bar \delta$
small enough.

\item $\sup_{t\in \mathbb{R}}\big|q_{h}(t)-\hat{q}_{h}(t)\big|<\frac{3L}{%
\sigma ^{2}R_{3}^{a-1}}+3R_{3}\varepsilon +\hat{R}(\varepsilon )h^{\beta }$,
for any $R_{3}\geq R_{2}$, and for all $h$ small enough.
\end{enumerate}
\end{lemma}

\textbf{Proof.}

\noindent\emph{Step1.} We prove the boundedness of the concentration point
$\hat q_h(t)$.

By the Theorem \ref{teoconcinf}, with $\varepsilon<1/2$, and by Remark
\ref{rembar4}, it is obvious that the ball $B(\hat q_h(t),\hat
R(\varepsilon)h^\beta)$ is not contained in the set
$\mathbb{R}^N\smallsetminus B(0,R_2)$, and we have
\begin{equation}
B(\hat q_h(t),\hat R(\varepsilon)h^\beta)\subset B(0,R_2+2\hat
R(\varepsilon)h^\beta).
\end{equation}
Because $\hat R(\varepsilon)$ does not depend on $h$, we can assume $h$ so
small that $2\hat R(\varepsilon)h^\beta<1$. Then
\begin{eqnarray}
&&|\hat q_h(t)|<R_2+2\hat R(\varepsilon)h^\beta<R_2+1; \\
&&B(\hat q_h(t),\hat R(\varepsilon)h^\beta)\subset B(0,R_2+1);
\label{form8bar}
\end{eqnarray}
This concludes the proof of the first claim.

\noindent\emph{Step 2.} We estimate the difference between the barycenter
and the concentration point.

We have
\begin{equation}
\big|q_{h}(t)-\hat{q}_{h}(t)\big|=\frac{\displaystyle\left\vert
\int_{\mathbb{R}^{N}}\big(x-\hat{q}_{h}(t)\big)u_{h}^{2}(t,x)dx\right\vert }
{\int_{\mathbb{R}^{N}}u_{h}^{2}(t,x)dx}
\end{equation}
and we split the integral in three parts, with $R_{3}\geq R_{2}$:
\begin{eqnarray*}
&&I_{1}=\frac{\left\vert \int\limits_{\mathbb{R}^{N}\smallsetminus B(0,R_{3})}
\big(x-\hat{q}_{h}(t)\big)u_{h}^{2}(t,x)dx\right\vert }
{\int\limits_{\mathbb{R}^{N}}u_{h}^{2}(t,x)dx}; \\
&&I_{2}=\frac{\left\vert \int\limits_{A_{2}}\big(x-\hat{q}_{h}(t)\big)
u_{h}^{2}(t,x)dx\right\vert }{\int\limits_{\mathbb{R}^{N}}u_{h}^{2}(t,x)dx}\
\text{ where }A_{2}=B(0,R_{3})\smallsetminus B(\hat{q}_{h}(t),\hat{R}(\varepsilon)h^\beta); \\
&&I_{3}=\frac{\left\vert \int\limits_{A_{3}}\big(x-\hat{q}_{h}(t)\big)
u_{h}^{2}(t,x)dx\right\vert }{\int\limits_{\mathbb{R}^{N}}u_{h}^{2}(t,x)dx}\
\text{ where }A_{3}=B(0,R_{3})\cap B(\hat{q}_{h}(t),\hat{R}(\varepsilon )h^\beta).
\end{eqnarray*}
It's trivial that $I_{3}\leq \hat{R}(\varepsilon )h^{\beta }$. By Lemma
\ref{lemmabar5} and by Theorem \ref{teoconcinf} we have
\begin{equation}
I_{2}\leq [2R_{3}+1]\varepsilon .
\end{equation}
By Step 1 and Remark \ref{rembar4} we have
\begin{equation}
|\hat{q}_{h}(t)|\frac{\int_{\mathbb{R}^{N}\smallsetminus
B(0,R_{3})}u_{h}^{2}(t,x)dx}{\int_{\mathbb{R}^{N}}u_{h}^{2}(t,x)dx}<(R_{3}+1)
\frac{L}{\sigma ^{2}R_{3}^{a}}<\frac{2L}{\sigma ^{2}R_{3}^{a-1}}.
\end{equation}
Also, by Remark \ref{rembar2}
\begin{equation}
\frac{\int_{\mathbb{R}^{N}\smallsetminus B(0,R_{3})}|x|u_{h}^{2}(t,x)dx}
{\int_{\mathbb{R}^{N}}u_{h}^{2}(t,x)dx}\leq \frac{L}{\sigma ^{2}R_{3}^{a-1}},
\end{equation}
hence
\begin{equation}
I_{1}\leq \frac{3L}{\sigma ^{2}R_{3}^{a-1}}.
\end{equation}
Concluding, we have that
\begin{equation}
\big|q_{h}(t)-\hat{q}_{h}(t)\big|<\frac{3L}{\sigma ^{2}R_{3}^{a-1}}
+3R_{3}\varepsilon +\hat{R}(\varepsilon )h^{\beta },
\end{equation}
for all $t\in \mathbb{R}$.

$\square $%

We notice that $R_1,R_2$ and $R_3$ defined in this section do not depend on
$\varepsilon$.

\subsection{Equation of the travelling soliton}

We prove that the barycenter dynamics is approximatively that of a point
particle moving under the effect of an external potential $V(x)$.

\begin{theorem}
\label{mainteoloc} Assume that $V$ satisfies (\ref{V0}),(\ref{Vinf*}),
(\ref{Vinf1}). Given $K>0,\ q\in \mathbb{R}^{N}$, let $\psi (t,x)\in
C(\mathbb{R},H^{2})\cap C^{1}(\mathbb{R},H^{1})$ be a global solution of equation
(\ref{schr}), with initial data in $B_{h}^{K,q}$, $h<h_{0}$. Then we have
\begin{equation}
\ddot{q_{h}}(t)+\nabla V(q_{h}(t))=H_{h}(t)
\end{equation}
with $||H_{h}(t)||_{L^{\infty }}$ goes to zero when $h$ goes to zero.
\end{theorem}

%\begin{proof}
\textbf{Proof.} We know by Theorem \ref{din}, that
\begin{equation}
\ddot{q_{h}}(t)+\frac{\int_{\mathbb{R}^{N}}\nabla V(x)u_{h}^{2}(t,x)dx}
{\int_{\mathbb{R}^{N}}u_{h}^{2}(t,x)dx}=0  \label{eqtrav1}
\end{equation}
Hence we have to estimate the function
\begin{equation}
H_{h}(t)=[\nabla V(\hat{q}_{h}(t))-\nabla V(q_{h}(t))]+\frac{\int_{\mathbb{R}^{N}}
[\nabla V(x)-\nabla V(\hat{q}_{h}(t))]u_{h}^{2}(t,x)dx}{\int_{\mathbb{R}^{N}}
u_{h}^{2}(t,x)dx}.  \label{eqtrav2}
\end{equation}
We set
\begin{equation}
M=\max_{{\tiny
\begin{array}{c}
\alpha =1,2 \\
|\tau |\leq K_{1}+R_{2}+1
\end{array}
}}|\partial ^{\alpha }V(\tau )|
\end{equation}
where $K_{1}$ is defined in Lemma \ref{lemmabar3} and $R_{2}$ is defined in
Remark \ref{rembar4}.

By Lemma \ref{lemmabar3} and Lemma \ref{lemmabar5} we get
\begin{eqnarray}
\big|\nabla V(\hat{q}_{h}(t))-\nabla V(q_{h}(t))\big| &\leq &\max_{{\tiny
\begin{array}{l}
i,j=1,\dots ,N \\
|\tau |\leq K_{1}+R_{2}+1
\end{array}
}}\left\vert \frac{\partial ^{2}V(\tau )}{\partial x_{i}\partial x_{j}}%
\right\vert |\hat{q}_{h}(t)-q_{h}(t)|\leq  \notag \\
&\leq &M\left[ \frac{3L}{\sigma ^{2}R_{3}^{a-1}}+3R_{3}\varepsilon +\hat{R}%
(\varepsilon )h^{\beta }\right] ,  \label{formH1}
\end{eqnarray}
for any $R_{3}\geq R_{2}$.

To estimate $$\frac{\displaystyle\int_{\mathbb{R}^N} [\nabla V(x)-\nabla
V(\hat q_h(t))]u_h^2(t,x)dx} {\displaystyle\int_{\mathbb{R}^N} u_h^2(t,x)dx}$$
we split the integral three parts.
\begin{eqnarray*}
L_1&=&\frac{ \displaystyle\int_{B(\hat q_h(t),\hat R(\varepsilon)h^\beta)}
|\nabla V(x)-\nabla V(\hat q_h(t))| u_h^2(t,x)dx}
{\displaystyle\int_{\mathbb{R}^N} u_h^2(t,x)dx}; \\
L_2&=&\frac{ \displaystyle\int_{\mathbb{R}^N\smallsetminus B(\hat
q_h(t),\hat R(\varepsilon)h^\beta)} |\nabla V(x)| u_h^2(t,x)dx}
{\displaystyle\int_{\mathbb{R}^N} u_h^2(t,x)dx}; \\
L_3&=&\frac{\displaystyle \int_{\mathbb{R}^N\smallsetminus B(\hat
q_h(t),\hat R(\varepsilon)h^\beta)}|\nabla V(\hat q_h(t))| u_h^2(t,x)dx}
{\displaystyle\int_{\mathbb{R}^N} u_h^2(t,x)dx}. \\
\end{eqnarray*}

By the Theorem \ref{teoconcinf} and by Lemma \ref{lemmabar5} we have
$L_3<M\varepsilon$.

By (\ref{form8bar}) and Lemma \ref{lemmabar5} we have
\begin{equation*}
L_{1}\leq \frac{\displaystyle\int\limits_{B(\hat{q}_{h}(t),\hat{R}
(\varepsilon )h^{\beta })}\max_{{\tiny
\begin{array}{l}
i,j=1,\dots ,N \\
|\tau |\leq K_{1}+R_{2}+1
\end{array}
}}\left\vert \frac{\partial ^{2}V(\tau )}{\partial x_{i}\partial x_{j}}%
\right\vert \hat{R}(\varepsilon )h^{\beta }u_{h}^{2}(t,x)dx}{\displaystyle%
\int_{\mathbb{R}^{N}}u_{h}^{2}(t,x)dx}\leq M\hat{R}(\varepsilon )h^{\beta }.
\end{equation*}
Now, using hypothesis (\ref{Vinf*}), equation (\ref{form8bar}), Theorem
\ref{teoconcinf} and Remark \ref{rembar2}, we have

\begin{eqnarray}
\nonumber&&\int\limits_{B(0,R_{2}+1)^C}
\big|\nabla V(x)\big| \frac{u_{h}^{2}(t,x)}{||u_{h}(t,\cdot )||_{L^{2}}^{2}}\leq\\
\nonumber&&\leq
\int\limits_{B(0,R_{2}+1)^C}\left[
\big|\nabla V(x)\big| \left(\frac{u_{h}^{2}(t,x)}{||u_{h}(t,\cdot )||_{L^{2}}^{2}}\right)^b
\left(\frac{u_{h}^{2}(t,x)}{||u_{h}(t,\cdot )||_{L^{2}}^{2}}\right)^{1-b}\right]
\\
\nonumber&&\leq \left[ \int\limits_{B(0,R_{2}+1)^C}\left[ \big|\nabla V(x)\big|\left(
\frac{u_{h}^{2}(t,x)}{||u_{h}(t,\cdot
)||_{L^{2}}^{2}}\right) ^{b}\right] ^{\frac{1}{b}}\right] ^{b}
\left[ \int\limits_{B(0,R_{2}+1)^C} \frac{u_{h}^{2}(t,x)}
{||u_{h}(t,\cdot)||_{L^{2}}^{2}}\right]
^{1-b}\\
\nonumber&&\leq \left[ \int\limits_{B(0,R_{2}+1)^C}\big| \nabla
V(x)\big|^{\frac{1}{b}}\frac{u_{h}^{2}(t,x)}{||u_{h}(t,\cdot
)||_{L^{2}}^{2}}dx\right] ^{b}\varepsilon ^{1-b}  \\
&&\leq \left[ \int\limits_{\mathbb{R}^{N}}V(x)\frac{u_{h}^{2}(t,x)}
{||u_{h}(t,\cdot )||_{L^{2}}^{2}}dx\right] ^{b}\varepsilon
^{1-b}\leq \left[ \frac{2L}{\sigma ^{2}}\right] ^{b}\varepsilon
^{1-b}\label{formL1-3}
\end{eqnarray}
where $b\in \left( 0,1\right)$ is defined in (\ref{Vinf*}). Furthermore,
again by Theorem \ref{teoconcinf} we have
\begin{equation}
\int\limits_{B(0,R_{2}+1)\smallsetminus B(\hat{q}_{h}(t),\hat{R}(\varepsilon
)h)}\big|\nabla V(x)\big|\frac{u_{h}^{2}(t,x)}{||u_{h}(t,\cdot
)||_{L^{2}}^{2}}dx\leq M\varepsilon .  \label{formL1-4}
\end{equation}
So, by (\ref{formL1-3}) and (\ref{formL1-4}),
\begin{equation}
L_{2}\leq M\varepsilon +\left[ \frac{L}{\sigma ^{2}}\right] ^{b}\varepsilon
^{1-b}.
\end{equation}
Concluding
\begin{equation}
\frac{\left\vert \int_{\mathbb{R}^{N}}[\nabla V(x)-\nabla V(\hat{q}%
_{h}(t))]u_{h}^{2}(t,x)dx\right\vert }{\int_{\mathbb{R}^{N}}u_{h}^{2}(t,x)dx}%
\leq 2M\varepsilon +\left[ \frac{L}{\sigma ^{2}}\right] ^{b}\varepsilon
^{1-b}+M\hat{R}(\varepsilon )h^{\beta }.  \label{formH2}
\end{equation}
Finally, by (\ref{formH1}) and (\ref{formH2}) we have
\begin{equation}
|H_{h}(t)|\leq \frac{3LM}{\sigma ^{2}R_{3}^{a-1}}+\left[ \frac{L}{\sigma ^{2}%
}\right] ^{b}\varepsilon ^{1-b}+M(2+3R_{3})\varepsilon +2M\hat{R}%
(\varepsilon )h^{\beta }.
\end{equation}
At this point we can have $\displaystyle\sup_{t}|H_{h}(t)|$ arbitrarily
small choosing firstly $R_{3}$ sufficiently large, secondly $\varepsilon $
sufficiently small, and finally $h$ small enough.

$\square $%\end{proof}

\textbf{Proof of Theorem \ref{teo1}.} By Theorem \ref{mainteoloc}
we get immediately the proof of Theorem \ref{teo1}.

$\square $

\end{document}